\documentstyle[eqsecnum,aps]{revtex}

\preprint{ZIMP-95-09}
\begin{document}
\title{
Quantum theory for mesoscopic electric circuits \\
}
\author{
You-Quan Li$^{\dag\ddag}$ and Bin Chen$^{\ddag}$
}
\address{
$^{\dag}$
CCAST(World Laboratory), P.O. Box 8730 Beijing 100080, P.R. China \\
\footnote{Mailing address}$^{\ddag}$
Zhejiang Institute of Modern Physics, Zhejiang University, Hangzhou
310027, P.R.China }
\date{\today}
\maketitle
\begin{abstract}
A quantum theory for mesoscopic electric circuits in accord with  the
discreteness of electric charges is proposed. On the basis of the theory,
Schr\"{o}dinger equation for
the  quantum LC-design and L-design is solved exactly. The uncertainty
relation for electric charge and current is obtained and a minimum
uncertainty state is solved. By introducing a gauge field, a formula for
persistent current arising from magnetic flux is obtained from a new point
of view.
\end{abstract}

\section{Introduction}

Along with the dramatic achievement in nanotechnology,
such as molecular beam epitaxy, atomic scale fabrication or advanced
lithography, mesoscopic physics and nanoelectronics are
undergoing a rapid development\cite{IBM,Buot}.
It has been a strong and definite trend in the miniaturization
of integrated circuits and components towards atomic-scale
dimensions\cite{Garcia} for the electronic device community.
When transport dimension reaches a characteristic dimension,
namely, the charge carrier inelastic coherence length, one must address
not only  quantum mechanical property but also
the discreteness of electron charge. Thus a correct quantum
theory is indispensable for the device physics in integrated circuits of 
nanoelectronics. Since the classical equation of motion for an electric
circuit of LC-design is just the same as that for
a harmonic oscillator whereas the `coordinate' has the meaning of electric
charge. the quantization of the circuit was carried out
\cite{Louisell} in the same way as that of a harmonic oscillator.
This only results energy quantization.
In fact, a different kind of fluctuation in
mesoscopic system, which inherently has nothing to do with energy
quantization and interference
of wave functions, is due to the quantization of electronic charge. 
Recently we studied the quantization of electric circuit of LC-design
under consideration of the discreteness of electric charge\cite{LiCh1}.

In present paper we extend the main idea of our previous letter\cite{LiCh1}
and present a quantum mechanical theory
for electric circuits based on the fact that electronic charge
takes discrete values.
In section \ref{sec:b},
A finite-difference Schr\'{o}dinger equation for the mesoscopic electric
circuit is obtained.
In section \ref{sec:c},
the Schr\"{o}dinger equation for a mesoscopic
circuit of LC-design is turned to  Mathieu equation in
`p-representation' and solved exactly. The average value of electric current
for ground state is calculated.
In section \ref{sec:d},
the uncertainty relation for charge and current
is discussed. a minimum uncertainty state, which recovers the usual
Gaussian wave packet in the limit of  vanishing of discreteness, is
solved.
In section \ref{sec:e}, the Schr\"{o}dinger equation for a quantum L-design
both in the presence of an adiabatic power source and
in the absence of source are solved exactly. A gauge field
is introduced and a formula for persistent current which is a periodic
function of the magnetic flux is obtained. It provides a formulation of
the persistent current in mesoscopic ring from a new point of view.
Finally, some discussions and conclusions are made in section \ref{sec:f}.

\section{ Quantization of electric circuit in accord with the
discreteness of electric charge }

We recall that for a classical  non-dissipative electric circuit of LC-design
in the presence of a source
$\varepsilon (t) $,
the equation of motion, as a consequence of Kirchoff's law, reads
$
{\displaystyle
\frac{d^2 q}{d t^2 } + \frac{1}{ LC } q - \frac{1}{ L } \varepsilon(t) = 0
}
$, 
where $q(t)$ stands for electric charge; L for inductance and C for
the capacity of the circuit. This equation of motion can
be formulated in terms of Hamiltonian mechanics, namely
\[
\dot{q} = \frac{\partial H }{\partial p }, \,\,
\dot{p} = - \frac{\partial H }{ \partial q }
\]
with
$
H(t) = {\displaystyle
\frac{ 1 }{ 2L } p^2 + \frac{ 1 }{2C}q^2 + \varepsilon(t) q
} $.
Here the variable $q$ stands for the electric charge instead of
the conventional `coordinate', while its conjugation variable
$
{\displaystyle
p(t) = L \frac{d q}{d t}
} $
represents (apart from a factor L ) the  electric current instead of
the conventional `momentum'. Analogous to the forced harmonic oscillator,
the electric circuit was quantized
by  many authors \cite{Louisell}, where the electric charge was treated
as a continuous variable. As a matter of fact, the electronic
charge is discrete and it must play an important role in the theory
for mesoscopic
circuits. Taking account of the discreteness of electric
charge, we must reconsider the quantization of a mesoscopic circuit.
According to the standard quantization
principle, one associate with each of the two observable quantities 
$q$ and $p$ a linear Hermitian operator, namely $\hat{q}$ and $\hat{p}$.
The Hamiltonian, also an observable quantity, corresponds to a Hermitian
operator
$\displaystyle{
H = \frac{1}{2L} \hat{p}^2 + V(\hat{q})
}$
which is a function of the operator $\hat{p}$ and $\hat{q}$. The 
commutation relation for the conjugation variables are
\begin{equation}
[ \hat{q}, \hat{p} ] = i \hbar
\label{eq:b}\end{equation}

Up to now, the discreteness of electronic charge is not
taken into account. Regarding to the discreteness, we must impose that
the eigenvalues of the self-adjoint operator $\hat{q}$ take discrete
values\cite{LiCh1}, i.e.
\begin{equation}
\hat{q} | q > = n q_e | q >
\label{eq:c}\end{equation}
where $n \in {\sf \, Z \!\!\! Z \, } $ (set of integers) and
$q_e  = 1.602 \times 10^{-19}$ coulomb,
the elementary electric charge.  Obviously, any eigenstate of
$\hat{q}$ can be specified by an integer. This allows us to
introduce a minimum `shift operator'
${\displaystyle
\hat{Q} := e^{iq_e \hat{p} /\hbar }
}$,
which is shown to satisfy the  following commutation relations\cite{LiCh1}
\begin{eqnarray}
[ \hat{q}, \hat{Q} ] = - q_e \hat{Q} \nonumber \\[4mm]
[ \hat{q}, \hat{Q}^{+} ] = q_e \hat{Q}^{+} \nonumber \\[4mm]
\hat{Q}^{+} \hat{Q} = \hat{Q} \hat{Q}^{+} = 1.
\label{eq:d}\end{eqnarray}
 
These relations can determine the structure of the whole Fock space. For
$
\hat{q} | n > = n q_e | n >
$, the algebraic relations (\ref{eq:d}) enable us to derive the followings
\begin{eqnarray}
\hat{Q}^{+} | n > = e^{i\alpha_{n+1} } | n + 1 >,
\nonumber \\
\hat{Q}| n > = e^{ -i\alpha_{n} } | n - 1 >,
\label{eq:qq}
\end{eqnarray}
where $\alpha_n $'s are undetermined phases. Obviously $\hat{Q}^+ $
and $\hat{Q}$ are ladder operators respectively for charge increasing and
decreasing in the diagonal representation of charge operator.
The Fock space for our present
algebra differs from the well known Fock space for the Heisenberg-Weyl
algebra, because the spectrum of the former is isomorphic to the set of
integers
${\sf \, Z \!\!\! Z \, }$
but that of the later is isomorphic to the set of non-negative
integers $ {\sf \, Z \!\!\! Z \, }^{+} + \{ 0 \} $.
Since 
$
\{ |n > | n \in {\sf \, Z \!\!\! Z \, } \}
$ constitute a Hilbert space, we have  the completeness
$
\sum_{n \in {\sf \, Z \!\!\! Z \, } } | n> < n | = 1
$. We also have the orthogonality
$
< n | m > = \delta_{n m}
$ due to the self-adjointness of $\hat{q}$.
As a result, the inner product in charge representation takes as
\begin{equation}
< \phi | \psi > 
= \sum_{n \in {\sf \, Z \!\!\! Z \, } } <\phi | n >< n |\psi >
=\sum_{n \in {\sf \, Z \!\!\! Z \, } } \phi^{*} (n) \psi (n)
\label{eq:e}\end{equation}

One can now study the eigenstates and eigenvalues of the operator
$\hat{p}$. Obviously, if
$ \hat{p} | p > = p | p > $
then
$ f(\hat{p}) | p > = f(p) | p > $
for any analytical function $f$.
Supposing
$ | p > = \sum_{n \in {\sf \, Z \!\!\! Z \, }  } c_n (p) | n > $,
and using
$ \hat{Q} | p > = e^{iq_e p /\hbar } | p > $ 
we can find that 
$ c_{n+1}/c_n = \exp(iq_e p / \hbar +i\alpha_{n+1} ) $,
which yields the following solution
\begin{equation}
| p > = \sum_{n \in {\sf \, Z \!\!\! Z \, } }
\kappa_n e^{i n q_e p/\hbar} | n >
\label{eq:f}\end{equation}
where
$ \kappa_n = e^{i\sum_{j=1}^{n}\alpha_j }$,
$ \kappa_{-n} = e^{-i\sum_{j=0}^{n-1}\alpha_{-j} }$ for $n > 0 $.
Obviously
$ |p + \hbar(2\pi /q_e ) > = | p >$, 
the eigenvalues of the operator $\hat{p}$ is a periodic parameter.
Topologically, the parameter space of the spectrum is isotopic to the $S^1 $.

Since the spectrum of charge is discrete and the inner product in charge
representation is a sum instead of the usual integral, one may define a
right and left discrete derivative operators  $\nabla_{q_e}$ and
$\overline{\nabla}_{q_e }$  by
\begin{eqnarray}
\nabla_{q_e} f(n) = \frac{ f(n+1) - f(n) }{ q_e }   \nonumber \\
\overline{\nabla}_{q_e } f(n) = \frac{f(n) - f(n-1) }{ q_e} .
\label{eq:g}\end{eqnarray}
They can be understood as the inverse of a discrete definite integral,
which is  in accord  with the inner product (\ref{eq:e}), i.e.
\begin{eqnarray*}
\int_{x_i }^{x_f }
f(x)dx & := &\sum_{n = n_i }^{n_f } q_e f( nq_e )\\
 \,    & =  & \left\{
\begin{array}{ll}
\hat{Q}F(x_f )  - F(x_i ) & if\,  \nabla_{q_e}F = f, \\
F(x_f ) - \hat{Q^+}F( x_i )     & if\, \overline{\nabla }_{q_e} F = f.
\end{array}  \right.
\end{eqnarray*}
Clearly, it recovers the conventional differential-integral calculus
as long as the minimum interval $ q_e $ goes to zero.
The discrete derivative operators defined by (\ref{eq:g}) can be expressed
explicitly by the minimum shift operators
\begin{eqnarray}
\nabla_{q_e} & = & (\hat{Q} - 1)/q_e \nonumber \\
\overline{\nabla}_{q_e } & = & ( 1 - \hat{Q}^+ ) /q_e .
\label{eq:gg}\end{eqnarray}
It is easy to check \cite{Li} that
$ \nabla_{q_e }^{+} = - \overline{\nabla }_{q_e } $.
Then we can write down two important
self-adjoint operators: `momentum' operator
\begin{equation}
\hat{P}= \frac{\hbar}{2i}
 ( \nabla_{q_e } + \overline{\nabla}_{q_e } )
= \frac{\hbar}{2iq_e } ( \hat{Q} - \hat{Q}^+ )
\label{eq:cp}\end{equation}
and free Hamiltonian operator
\begin{equation}
\hat{H}_0 =
 - \frac{\hbar^2}{2}
\nabla_{q_e}\overline{\nabla}_{q_e }
= - \frac{\hbar^2}{2q_e}
( \nabla_{q_e } - \overline{\nabla }_{q_e } ) ,
= - \frac{\hbar^2 }{ 2 q_e^2 } (\hat{Q} + \hat{Q}^+  - 2 )
\label{eq:fh}
\end{equation}
where we call them respectively as momentum and free Hamiltonian operators
because they are really those when $q_e \rightarrow 0$.
Now we have finished the
quantization of mesoscopic electric circuits  and obtained the following
finite-difference Schr\"{o}dinger equation,
\begin{equation}
\left[
- \frac{\hbar^2 }{ 2q_e L} (\nabla_{q_e } -\overline{\nabla}_{q_e } )
+ V(\hat{q})
\right] | \psi >  = E | \psi >.
\label{eq:h}\end{equation}

\label{sec:b}

\section{The quantum LC-design}

As an application of our quantization strategy of the mesoscopic
circuit, we discuss a mesoscopic LC-design in this section.
We only consider the adiabatic approximation so that
$\varepsilon (t)$ is consider as a constant $\varepsilon$. Then the
Schr\"{o}dinger equation (\ref{eq:h}) for a LC-design is written as
\begin{equation}
\left[- \frac{\hbar^2 }{ 2q_e L} (\nabla_{q_e } -\overline{\nabla}_{q_e } )
+ \frac{1}{ 2C} \hat{q}^2 + \varepsilon \hat{q}
\right] | \psi > = E  |\psi >
\label{eq:lch}\end{equation}

We consider a representation in which the operator $\hat{p}$ is
diagonal and called it as p-representation. We must address that
the $\hat{p}$ is the conjugation of the charge variable $\hat{q}$ within
the meaning of  usual canonical commutator (\ref{eq:b}), and it is
the `current' operator only if the charge is treated as a continuous
variable. However, the operator $\hat{P}$ associated with the
physical quantity, electric current(apart from a factor
$\displaystyle\frac{1}{L}$),
differs from the operator $\hat{p}$ as long as the discreteness of charge
is taken into account. Clearly, $\hat{P}$ will become the usual $\hat{p}$
when $q_e $ goes to zero. The orthogonality of eigenstates of
$\hat{p}$ is an immediate consequence of (\ref{eq:f}) and the orthogonality
of the charge eigenstates, i.e.
$< p | p' > =\displaystyle \frac{ 2\pi }{ q_e \hbar }
\sum_{n\in{\sf \, Z \!\!\! Z \, }}\delta(p-p'+n(
\frac{2\pi}{q_e})\hbar ).
$
The completeness is also verified
\begin{equation}
\frac{q_e}{2\pi}\int^{\hbar(\frac{\pi}{q_e} ) }_{-\hbar(\frac{\pi}{q_e}) }
\frac{dp}{\hbar}|p >< p| = \sum_{n \in{\sf \, Z \!\!\! Z \, } } |n >< n| = 1.
\label{eq:i}\end{equation}
The transformation of wave functions between charge representation
and p-representation is given by
\begin{equation}
< n | \psi > = (\frac{q_e}{2\pi\hbar} )
\int^{\hbar(\frac{\pi}{q_e}) }_{-\hbar(\frac{\pi}{q_e} ) }
dp <p |\psi >
e^{
-in\frac{q_e p }{\hbar}
}
\label{eq:j}\end{equation}

Using (\ref{eq:f}), we can obtain the following relations
\begin{eqnarray}
< p' | \nabla_{q_e} - \overline{\nabla}_{q_e} | p >
& = & \frac{4\pi\hbar}{ q_{e}^2 }
\left( \cos(\frac{q_e}{\hbar}p ) - 1 \right)\delta(p - p') \nonumber \\
<p' | \,  \hat{q}^2  \,| p >
& = & -\frac{2\pi \hbar^3 }{q_e}
\frac{\partial ^2 }{\partial p^2 }
\delta (p - p')
\label{eq:k}\end{eqnarray}
In the `p-representation', the finite-difference Schr\"{o}dinger
equation (\ref{eq:lch}) becomes a differential equation for
$
\tilde{\psi }(p) := < p | \psi >
$
\begin{equation}
\left[
- \frac{\hbar^2 }{2C} \frac{\partial^2 }{\partial p^2 }
- \frac{\hbar^2}{q_{e}^2 L }( \cos(\frac{q_e}{\hbar} p ) - 1 )
\right] \tilde{\psi } (p) = E\tilde{\psi } (p).
\label{eq:l}\end{equation}
which is the well known Mathieu equation \cite{Wang,Grad}. This equation
was ever appeared in \cite{Jurk} on the discussion of Pad\'{e} approximates.
In deriving to (\ref{eq:l}),
we have adopted $\varepsilon = 0 $ for simplicity.
Actually, the linear term in (\ref{eq:lch}) can be moved by a
translation in the `coordinate' (charge) space.
Apart from  a re-definition of
$ \hat{q} $ and a shift of the energy $E$, the same equation as (\ref{eq:l})
would be derived.

In terms of the conventional notations \cite{Wang,Grad}, the wave
functions in p-representation can be solved as follows
$$
\tilde{\psi}^{+}_{l}(p) =
{\rm ce }_l (\frac{\pi}{2}- \frac{q_e }{ 2\hbar}p , \,\xi )
$$
or
\begin{equation}
\tilde{\psi}^{-}_{l+1}(p) =
{\rm se }_{l+1} (\frac{\pi}{2} - \frac{q_e }{2 \hbar}p ,  \,\xi )
\label{eq:m}\end{equation}
where the superscripts `+' and `-' specify  the even and odd parity
solutions respectively;
$l = 0, 1, 2, \cdots $;
$\xi = {\displaystyle (\frac{2\hbar}{q^{2}_e } )^2
\frac{C}{L} } $; ce$(z, \xi)$ and se$(z, \xi)$ are periodic Mathieu
functions.  In this case, there exist infinitely many eigenvalues
$\{ a_l \}$ and $\{ b_{l+1} \} $ which are not identically equal to zero.
Then the energy spectrum is expressed in terms of the
eigenvalues $a_l $, $b_{l}$ of Mathieu equation

\begin{eqnarray}
E^{+}_l =  \frac{q_{e}^2 }{ 8 C }a_l (\xi)
+ \frac{\hbar^2 }{ q_{e}^2 L }
\nonumber \\
E^{-}_{l+1} =  \frac{q_{e}^2 }{ 8 C }b_{l+1}(\xi)
+ \frac{\hbar^2 }{ q_{e}^2 L }
\label{eq:n}\end{eqnarray}

As an exercise, one may calculated the fluctuation of electric current
for the ground state. It is known that the explicit results of eigenvalues
and eigenfunctions of Mathieu equation are complicated. They are related
to continue fractions and trigonometric series respectively.
For the concrete values of the Plank constant and the elementary
electric charge, the WKB method is valid. From the series solution of
Mathieu equation for ground state, we obtained the fluctuation of
electric current $\hat{P}$ (apart from a factor $1/L$) for ground state,
\begin{equation}
<  \hat{P}^2 >_{\rm ground} =
\frac{1}{2} \left( \frac{h}{q_e}
            \right)^2
\left[ 1 - \frac{3}{2}\left( \frac{\hbar^2 C}{q_{E}^4 L }
                      \right)^2 + \cdots
\right].
\end{equation}
This result is valid for the case
$C/L <<(q_e^2/\hbar )^2$.
\label{sec:c}

\section{Uncertainty relation and the minimum uncertainty state}

In order to fairly understand the main conclusions in this section, we begin
with a brief view of the derivation of the uncertainty relation in standard
quantum mechanics. If $\hat{A}$ and $\hat{B}$ are two Hermitian (self-adjoint)
operators which do not commute, the physical quantities $A$ and $B$ cannot
both be sharply defined simultaneously. The variances of $A$ and $B$ are
defined as $ (\Delta \hat{A})^2 = < (\hat{A} - < \hat{A} > )^2 > $ and
$ (\Delta \hat{B})^2 = < (\hat{B} - < \hat{B} > )^2 > $.
Their positive square roots, $\Delta A$ and $\Delta B$ are called the
uncertainties in $A$ and $B$. In terms of the properties of self-adjoint
operators and the knowledge of Schwarz inequality, one can prove that
\begin{equation}
(\Delta\hat{A})^2 (\Delta\hat{B})^2
\geq \, \mid < \frac{1}{2} ( \{ \hat{A}, \hat{B} \} - <A><B> ) > \mid^2
+ \mid < \frac{1}{2} [ \hat{A}, \hat{B} ] > \mid^2
\label{eq:da}
\end{equation}
where $\{ \, , \, \}$ denotes the anti-commutator, and the equality sign
holds if and only if
$ \hat{B}| \psi > \propto \hat{A} |\psi > $.
In deriving (\ref{eq:da}), the fact that the expectation value of a Hermitian
(or anti-Hermitian) operator is a real number (or purely imaginary number)
has been used. As a direct consequence of (\ref{eq:da}), the uncertainty
relation is conventionally written as
\begin{equation}
(\Delta\hat{A})^2 (\Delta\hat{B})^2
\geq \, \mid < \frac{1}{2} [ \hat{A}, \, \hat{B} ] > \mid^2 .
\label{eq:db}
\end{equation}
Clearly. the equality sign in (\ref{eq:db}) holds if and only if both the
equality sign in (\ref{eq:da}) holds and the first  term of the right
hand side in (\ref{eq:da}) vanishes. These conditions imply that
\begin{eqnarray}
( \hat{B} - < \hat{B} > ) | \psi >
& = & \lambda (\hat{A} - <\hat{A}> ) |\psi >
\nonumber \\
\lambda & = & \frac{ <\psi | [ \hat{A}, \, \hat{B} ] | \psi >}
{2 (\Delta \hat{A} )^2 }
\label{eq:dc}
\end{eqnarray}

Now we direct to our main purpose. After some calculations, we obtain the
following commutation relations for the charge $\hat{q}$, the current
$\hat{P}$ and the free Hamiltonian $\hat{H}_0 $
\begin{equation}
[ \hat{H}_0 , \, \hat{P} ] = 0, \,\,
[ \hat{H}_0 , \, \hat{q} ] = i \hbar \hat{P}, \,\,
[ \hat{q} , \, \hat{P} ] = i \hbar (1  + \frac{ q^2_e }{ \hbar^2 }\hat{H}_0 )
\label{eq:dd}
\end{equation}
where the operators $\hat{P}$ and $\hat{H}_0 $ have been defined
respectively by (\ref{eq:cp}) and (\ref{eq:fh}). The term
${ \displaystyle \frac{q^2_e }{\hbar^2 } \hat{H}_0 } $
in the third equation of (\ref{eq:dd})
occurs due to the discreteness of electric charge. Now we are ready to
write out the uncertainty relation for electric charge and electric current,
namely
\begin{equation}
\Delta\hat{q}\cdot\Delta\hat{P} \geq \frac{\hbar}{ 2 }
( 1 + \frac{q^2_e }{\hbar^2} < \hat{H}_0 > ).
\label{eq:de}
\end{equation}
This uncertainty relation recovers the usual Heisenberg
uncertainty relation if $ q_e $ goes to zero, i.e. the case that the
discreteness of electric charge vanishes. Moreover, the uncertainty relation
(\ref{eq:de}) has shown us some new sense beyond the knowledge of traditional
Heisenberg uncertainty relation.

It is of interest to study the particular state $ | \psi >$, for which
(\ref{eq:de}) becomes an equality. This is the state in which the product
of the uncertainties in electric charge and current is as small as the
noncommutivity allows:
$
\Delta\hat{q}\cdot\Delta\hat{P} =\displaystyle
\frac{\hbar}{ 2 }
( 1 +  \frac{q^2_e }{\hbar^2} < \hat{H}_0 > ).
$
Such a minimum uncertainty state must obey the condition (\ref{eq:dc}) for
$ \hat{A} = \hat{P} $ and $ \hat{B} = \hat{q} $
\begin{equation}
( \hat{q}\, - < \hat{q} > ) | \psi >= -
\frac{ i \hbar ( 1 + {\displaystyle\frac{q^2_e }{\hbar^2 }  }
< \hat{H}_0 > ) }
{ 2 ( \Delta\hat{P} )^2 }
(\hat{P}\, - < \hat{P} > ) | \psi >
\label{eq:df}
\end{equation}
Using (\ref{eq:f}) and (\ref{eq:qq}), one can find that
\begin{eqnarray}
< p' |\, \hat{q} \, | p >
& = &  \frac{ h }{ q_e }\frac{\hbar}{i} \frac{\partial }{ \partial p }
\delta(p - p') \nonumber \\
<p' |\, \hat{P}\, | p >
& = & \frac{ h }{ q_e } \frac{\hbar}{ q_e } \sin(\frac{q_e p }{\hbar})
\delta (p - p')
\end{eqnarray}
Then (\ref{eq:df}) becomes the following differential equation in
p-representation
\begin{equation}
\left(
\frac{\hbar}{i}
\frac{\partial }{\partial p } \, + < \hat{q} >
\right) \tilde{\psi}(p) =
\frac{ i \hbar ( 1 + \displaystyle\frac{q^2_e }{\hbar^2 } < \hat{H}_0 > ) }
{ 2 ( \Delta\hat{P} )^2 }
\left(
\frac{\hbar}{q_e}\sin(\frac{q_e p }{\hbar})\, - < \hat{P} >
\right) \tilde{\psi } (p).
\label{eq:dg}
\end{equation}
This differential equation is solved by a plane wave with modulated
amplitude:
\begin{equation}
\tilde{\psi}(p) = N \exp
\left[
\frac{ 1 + {\displaystyle \frac{q^2_e }{\hbar^2 }  }< \hat{H}_0 > }
{ 2 ( \Delta\hat{P} )^2 }
\left( \, \frac{\hbar^2 }{ q^2_e } \cos (\frac{q_e}{\hbar}p)\, + <\hat{P}>p
\right)
- \frac{ i < \hat{q} > p }{ \hbar }
\right].
\label{eq:dh}
\end{equation}
where $N$ is the normalization constant. (\ref{eq:dh})
is obviously a deformation of the usual Gaussian wave packet
and recovers the Gaussian wave-packet if the discreteness vanishes.

\label{sec:d}

\section{Quantum L-Design, Gauge Field and Persistent Currents }
\label{sec:e}
In this section, we will solve the Schr\"{o}dinger equation for a L-design
in the presence of an adiabatic source and in the absence of source.
Introducing a gauge field and gauge transformation, we derive a formula
for persistent current in a pure L-design, i.e. a mesoscopic metal ring.

\subsection{ The L-design  in the presence of an adiabatic source}
The Schr\"{o}dinger equation for a L-design in the presence of an adiabatic
source reads
\begin{equation}
\left[- \frac{\hbar^2 }{ 2q_e L} (\nabla_{q_e } -\overline{\nabla}_{q_e } )
 + \varepsilon \hat{q}
\right] | \psi > = E  |\psi >.
\label{eq:lh}\end{equation}
In order that the quantization of a mesoscopic circuit be valid, the size
of the circuit must be restricted. While the voltage source can come from
an infinite reservoir to keep the chemical potential constant.
We consider present problem in charge representation and expand the eigenstate
of (\ref{eq:lh}) in terms of orthonormal set of charge eigenstates,  namely,
$ | \psi > = \sum^{\infty}_{n=-\infty} u_n \, | n > $.
Substituting it into (\ref{eq:lh}), we obtain the following recursion
relations
\begin{equation}
2( \frac{\hbar^2 }{q_e L } + n q_e \varepsilon  -E )u_l
- \frac{\hbar^2 }{ 2q_e L }( u_{l-1} + u_{l+1} ) = 0
\label{eq:eb}
\end{equation}

The knowledge of recursion formula of Bessel functions,
$
z ( J_{\nu+1} (z) + J_{\nu-1}(z) \, ) = 2\nu J_\nu (z)
$
enables us to write down a solution of (\ref{eq:eb})
\begin{equation}
u_n = J_{nq_e\varepsilon + z_0 - E } ( z_0 )
\label{eq:ec}
\end{equation}
where $ z_0 =\displaystyle\frac{\hbar^2 }{q_e L}$. In terms of (\ref{eq:ec})
the eigenstates of Schr\"{o}dinger equation (\ref{eq:lh}) are write out
\begin{equation}
| \psi_E > = \sum^{\infty}_{n= -\infty}
J_{nq_e\varepsilon + z_0 - E } ( z_0 ) | n >
\label{eq:ed}
\end{equation}
which is the solution  of eigenstates
for quantum L-design in presence of a adiabatic source.

\subsection{ Pure L-design }
Now we consider a pure L-design,
$
\hat{H}_L =
- {\displaystyle \frac{\hbar^2 }{ 2q_e L}  }
(\nabla_{q_e } -\overline{\nabla}_{q_e } )
$.
The Hamiltonian operator of the pure L-design $\hat{H}_L$,
the current operator $\hat{P}$ and the operator $\hat{p}$ commute each other, so they
can have simultaneous eigenstates. Actually, (\ref{eq:f}) is the simultaneous
eigenstate of those operators, i.e.
\begin{eqnarray}
\hat{P} | p > & = & \frac{\hbar}{q_e}\sin(\frac{q_e p}{\hbar})\,| p >,
\nonumber \\
\hat{H}_L | p > & = & \frac{\hbar^2}{q_e^2 L}
\left( 1 - \cos (\frac{q_e p}{\hbar} )\,
\right)\,| p >,
\label{eq:ef}
\end{eqnarray}
This result tells us that the magnitude of electric current in a mesoscopic
electric circuit of pure L design are bounded taking values between
$ - (\displaystyle\frac{\hbar}{q_e L}) $
and 
$ (\displaystyle\frac{\hbar}{q_e L}) $.
It also indicates that the maximum
quantum noise in a pure L-design ( a mesoscopic ring is an example)
takes finite value if the elementary charge 
$q_e $ should not be considered as the infinitesimal ( particularly for
the mesoscopic circuit).  It is also
worthwhile to notice that both the current and energy of a pure L-design
become null when
$ p =\displaystyle \frac{ 2\pi \hbar}{q_e}$ as long as $ q_e$ is not
zero. Clearly, the lowest energy states correspond to
$p = n \displaystyle\frac{h}{q_e}$
for any integer $n$. Thus  the energy spectrum is infinitely degenerated.

\subsection{ Gauge field and persistent current }

In previous discussion, we have used the terminology `p-representation'
and solved the eigenstates of $\hat{p}$. Now let us to find out
what the eigenvalues of $\hat{p}$
mean. If introducing a operator
$
\hat{G} := e^{ -i\beta\frac{\hat{q}}{\hbar} }
$, we can find that
$
\hat{G} | p > = | p - \beta >
$ and
$
\hat{G}^+ | p > = | p + \beta >
$.
Considering a unitary transformation to the eigenstates of
Schr\"{o}dinger operator given by
\[
| \psi > \rightarrow  | \psi' > = \hat{G} | \psi >,
\]
we find that the Schr\"{o}dinger equation (\ref{eq:h}) is not covariant.
This requests us to introduce a gauge field and to define a reasonable
covariant discrete derivative. By making the following definitions:
\begin{eqnarray}
D_{q_e} := e^{-\frac{q_e}{\hbar}\phi }
     \frac{ \hat{Q} - e^{ i\frac{ q_e}{\hbar} \phi } }
          { q_e } , \nonumber \\
\overline{D}_{q_e} := e^{\frac{q_e}{\hbar}\phi }
\frac{ e^{ -i\frac{ q_e}{\hbar} \phi } - \hat{Q^+}
}{ q_e } ,
\label{eq:eg}
\end{eqnarray}
we can verify that they are covariant under a gauge transformation. The
gauge transformations are expressed as
\begin{eqnarray}
\hat{G} D_{q_e} \hat{G}^{-1} = D'_{q_e}, \nonumber \\
\hat{G} \overline{D}_{q_e} \hat{G}^{-1} = \overline{D}'_{q_e},
\label{eq:eh}
\end{eqnarray}
as long as the gauge field $\phi$ transforms in such a way
\[
\phi \rightarrow \phi' = \phi - \beta.
\]
From either the transformation law or the dimension of the field $\phi $,
we may realize that $\phi$ plays the role of the magnetic flux threading
the circuit.

In terms of those covariant discrete derivatives (\ref{eq:eg}), one can
write down the Schr\"{o}dinger equation in the presence of the gauge field
(magnetic flux). Here we write out the Schr\"{o}dinger equation for
a pure L-design in the presence of magnetic flux,
\begin{equation}
-\frac{\hbar^2 }{2q_e L } (D_{q_e}\, - \overline{D}_{q_e} )
| \psi > = E | \psi >.
\label{eq:gh}
\end{equation}
Because its eigenstates can be simultaneous eigenstates of $\hat{p}$.
(\ref{eq:gh})  is solved by the same eigenstate $| p >$ in (\ref{eq:f}).
The energy spectrum is easily calculated  as
\begin{equation}
E(p,\phi) = \frac{2\hbar}{q^2_e} \sin^2
\left( \frac{q_e}{2\hbar}(p-\phi)
\right)
\end{equation}
which has oscillatory property with respect to
$\phi$ or $p$. Differing from the usual classical pure L-design, the
energy of a mesoscopic quantum pure L-design can not be large than
$2\hbar / q_e^2 $.
Clearly, the lowest energy states are those states that
$
p = \phi + n (\displaystyle\frac{h}{q_e})
$.
Thus the eigenvalues of the electric current ( i.e.
$ \displaystyle { \frac{1}{L}\hat{P}  }$ ) of ground state are calculated
\begin{equation}
I(\phi) = \frac{\hbar}{q_e L} \sin( \frac{q_e}{\hbar} \phi ).
\label{eq:pc}
\end{equation}
Obviously, the electric current on a mesoscopic  circuit of pure L-design
is not null in the presence of a magnetic flux except
$\phi = n\displaystyle(\frac{h}{q_e}$).
Clearly, this is a pure quantum characteristic.
(\ref{eq:pc})  exhibits that the persistent current in a mesoscopic L-design
is an observable quantity periodically depending on the flux $\phi$.
Because a mesoscopic metal ring is a natural pure L-design, the formula
(\ref{eq:pc}) is valid for persistent current on a single mesoscopic
ring\cite{Chand}. Differing from the conventional formulation of the persistent
current on the basis of quantum dynamics for electrons, our formulation
presented a  method from a new point of view. Formally, the $I(\phi)$
we obtained here is a sine function with periodicity of
$\phi_0 =\displaystyle\frac{h}{q_e}$, But either the model that the electrons move freely
in an ideal ring\cite{Ch1}, or the model that the electrons have hard-core
interactions between them\cite{LiMa2} can only give the sawtooth-type
periodicity. Obviously, the sawtooth-type function is only
the limit case for $q_e /\hbar \rightarrow 0 $.

Certainly  the experiment \cite{Mail}
should be considered as the case of persistent current in a LC-design
because the junction of semiconductors will contribute a capacitance
to the `circuit'.

\section{Conclusions and discussions }

In the above, we studied the quantization of  mesoscopic electric
circuit. Differing from the literature in which it is simply treated
as the quantization of a harmonic oscillator, we addressed the importance
of the discreteness of electric charge. Taking the discreteness into
account, we proposed a quantum theory for mesoscopic electric circuit
and give a finite-difference Schr\'{o}dinger equation for mesoscopic
electric circuit. As the Schr\"{o}dinger equation for LC-design in
p-representation becomes the well known Mathieu
equation, it is exactly solved. We obtain the wave functions
in terms of Mathieu functions and the energy
spectrum in terms of the eigenvalues of Mathieu equation.
The discussion on uncertainty relation for the charge and current shed some
new light on the knowledge of transitional Heisenberg uncertainty relation.
The discreteness of electric charge increased the uncertainty which is
related to the expectation value of the `free' Hamiltonian. The minimum
uncertainty state we obtained is a deformation of the standard Gaussian
wave packet. As further applications  of our theory, the eigenstates of
L-design in the presence and in the absence of source were solved
respectively. Introducing a gauge field and gauge transformation,
we successfully obtained a formula for the persistent current on
the mesoscopic pure L-design in the presence of the magnetic flux.
As the mesoscopic metal ring is a natural pure L-design,
the formula is certainly valid for the persistent current on
mesoscopic rings.
In our formula, the mass of electrons, the carriers for electric current,
is not involved. This is worthwhile to check by some experiment.
Our present theory is believed to explain the Coulomb blockade
on which the research is in progress.

In addition, all the results in present paper will recover  the standard
knowledge if one takes the continuous limit $q_e \rightarrow 0 $,
e.g. (\ref{eq:pc}) becomes $\phi = L I $ in the limit of
$q_e \rightarrow 0$,
the well known formula in electromagnetism.
So the whole theory and their results are believed to be consistent and
reasonable. One may noticed that we used the charge representation and
the so called p-representation. Because of the discreteness of electric
charge, $\hat{p}$ is no longer a current operator, but should be
understood as the usual Dirac conjugation of the charge operator
satisfying (\ref{eq:b}) only.
The operator $\hat{P}$ which is associated with physical
observable, electric current, obeys the commutation relation
(\ref{eq:dd}).
Clearly, the current operator $\hat{P}$ is not a Dirac conjugation
of the charge operator $\hat{q}$. So we need a new definition
about such conjugation defined by (\ref{eq:dd}).
\label{sec:f}
\section*{Acknowledgement}
The work is supported by NSFC and NSF of Zhejiang Province.

\end{document}